\documentclass[10pt,letterpaper]{article}
\usepackage[top=0.85in,left=2.75in,footskip=0.75in]{geometry}

\usepackage{amsmath,amssymb}

\usepackage{changepage}

\usepackage{textcomp,marvosym}

\usepackage{cite}

\usepackage{nameref,hyperref}

\usepackage{microtype}
\DisableLigatures[f]{encoding = *, family = * }

\usepackage[table]{xcolor}

\usepackage{array}

\newcolumntype{+}{!{\vrule width 2pt}}

\newlength\savedwidth

\usepackage{caption}
\usepackage{subcaption}
\usepackage{algorithm}
\usepackage{algpseudocode}
\usepackage{pdflscape}
\usepackage{gensymb}

\usepackage{color}

\raggedright
\usepackage[aboveskip=1pt,labelfont=bf,labelsep=period,justification=raggedright,singlelinecheck=off]{caption}

\makeatletter
\renewcommand{\@biblabel}[1]{\quad#1.}
\makeatother

\date{}

\usepackage{xcolor,cite,etoolbox}
\makeatletter 
\pretocmd\@bibitem{\color{black}\csname keycolor#1\endcsname}{}{\fail}
\newcommand\citecolor[1]{\@namedef{keycolor#1}{\color{blue}}}
\makeatother

\usepackage{lastpage,fancyhdr,graphicx}
\usepackage{epstopdf}
\fancyheadoffset[L]{2.25in}
\fancyfootoffset[L]{2.25in}
\begin{document}
\vspace*{0.2in}

%
%
%
%
%
%
%

\begin{flushleft}
{\Large
\textbf\newline{Analysing Human Mobility Patterns of Hiking Activities through Complex Network Theory} 
}
\newline
\\
Isaac Lera\textsuperscript{1*},
Toni P\'erez\textsuperscript{2},
Carlos Guerrero\textsuperscript{1},
V\'ictor M. Egu\'iluz\textsuperscript{2},
Carlos Juiz\textsuperscript{1}
\\
\bigskip
\textbf{1} Departamento de Matem\'aticas e Inform\'atica. Universitat de les Illes Balears, Palma de Mallorca, Spain
\\
\textbf{2} Instituto de F\'isica Interdisciplinar y Sistemas Complejos IFISC (CSIC-UIB), Palma de Mallorca, Spain
\\
\bigskip

%
%





* E-mail: isaac.lera@uib.es (IL)

\end{flushleft}

%

\section*{Abstract} 
The exploitation of high volume of geolocalized data from social sport tracking 
applications of outdoor activities can be useful for natural resource planning 
and to understand the human mobility patterns during leisure activities.  This 
geolocalized  data represents the selection of hike activities according to 
subjective and objective factors such as personal goals, personal abilities, 
trail conditions or weather conditions. In our approach, human mobility patterns are analysed from  
trajectories which  are generated by hikers. We propose the generation of the 
trail network identifying special points in the overlap of trajectories. Trail 
crossings and trailheads  define our network and shape topological features. 
We analyse the trail network of Balearic Islands, as a case of study, using 
complex weighted network theory. The analysis is divided into the four seasons 
of the year to observe the impact of weather conditions on the network topology. 
The number of visited places does not decrease despite the large difference
in the number of samples of the two seasons with larger and lower activity.
It is in summer season where it is produced the most significant variation in the frequency and localization of activities from inland regions to coastal areas.
Finally, we compare our model with other related studies where the network possesses a different purpose. 
One finding of our approach is the detection of regions with relevant importance where landscape interventions can be applied in function of the communities.


\section*{Introduction}
The increment of outdoor activities in recent years has had an economical and 
an environmental impact for a wide number of parties: landscape owners, 
environmental organizations, public administrations, and tourism entities among others. 
The extended use of GPS devices in outdoor activities together with the proliferation of social sport trackers sites enables people to share and comment their tracks and training measurements. This type of data sources provides large volume of information that can be analysed. For example, the occupation of outdoor areas or the characterization of human mobility patterns can be inferred from the records of social sport trackers \cite{Song2010,cortes2014} 
In addiction, understanding the factors that influence the practice of a certain sport has a wide range of applications: for example in preserving and planing natural areas~\cite{Orellana2012a,Beeco2014a,Meijles2014a,Beeco2013a}, determining the most frequented attractions~\cite{Paldino2015a,Orellana2012a,Yuan2012a,Huang2009a, Mulligann2011a}, predicting agglomerations~\cite{Giannotti2011,Biljecki2013a}, discovering travel patterns~\cite{Zheng2012,Yuan2012a,Brockmann2006a,Hoogendoorn2005}, or measuring the overlapping of activities~\cite{Beeco2014a}.

The problem of modelling human behaviour in urban and natural environments has 
been widely studied. In urban environments, human mobility data have been extracted from internet and mobile phone 
connections~\cite{Song2010,Schneider20130246,Montjoye13,gonzalez08}, 
geolocalized tweets~\cite{Bassolas2016,Jurdak2015,FriasMartinez2014a}, and GPS-tagged 
photos~\cite{Zheng2012,Arase2010,Kisilevich2010} among other sources. 
In natural environments, generally natural parks, visitors or hikers mobility 
data come from GPS devices instead of mobile phone connections due to 
insufficient precision or the lack of coverage. 
To obtain trajectories from visitors, in several cases, authors provide them 
preconfigured GPS devices to facilitate the applicability of statical 
techniques within small confined areas~\cite{Orellana2012a,Beeco2014a,Giannotti2007}. 
In the analysis of the movement, several techniques are used such as 
point density~\cite{Meijles2014a,Tracy,Arrowsmith2002295}, counting the 
frequency of visits on a determined area~\cite{Beeco2013a,Shoval2010}, 
identifying suspension patterns (areas with a reduction in the speed), and flow 
distribution among areas~\cite{Orellana2012a}. In these studies, the goal is to 
obtain a rank of areas or trail segments according to some characteristic 
(e.g. trail degradation, frequency, etc.).

We provide an approach for the use of geolocalized activity to build a network 
that allows the identification of places of relevance and the relationship among them. 
Other studies addresses the problem of finding points of interest by considering small specific 
and confined regions without exploring interferences with other environments and external factors. 
The environment choices can dictate the flow of the movement and the pressure on it forming a network of paths. 
In hiking activities, the structure of paths can be associated to roads through which hikers perform their activities 
reflecting the conditions of them: access, services, and other factors such as personal goals or weather conditions. 
The transformation of the activity into a network allows to the analysis and identification of places 
of relevance where interventions can be conducted.
We perform a complex network analysis of 15376 hiking routes performed during 2009-2016 in the Balearic Islands. 
Our dataset, coming from a sport tracker application, is highlighted by the variability of cases, by the 
topography of each island, and by the fact that the uploaded data are not conditioned by the 
study.  We conduct various topological analysis taking into account external 
factors such as the seasonal weather, the difficulty of the route, or 
trailhead facilities. 

Complex network theory enables the extraction of non singular topological 
features~\cite{Strogatz2001}. In the literature, there are several cases of applicability  
to transport system~\cite{Barthelemy2011,watts1998cds,Latora2002109, Poland22,Zhang2016,Soh2010,TIAN2016537}. 
For example, in~\cite{Soh2010} the transport network of Singapore subway has been analysed where the nodes 
are transport stops and stations, the edges represent the transport lines connection between the nodes, 
and the weights of the edges are the number of passengers between the nodes.  
Another case of study is the human impact in natural environments~\cite{Knights13}, 
where different types of ecological characteristics to describe the interaction between human activities and the ecosystem are combined. 
In our approach, the nodes of the network are head and cross trails, and the edges represent the hiking activity that join these elements. 
Thus, the modelling of the network provides the location of points of relevance and the relationship between them, 
as well as the flow of the movement and indicators about the pressure of certain regions.

From our approach, we derived the following research issues:
\begin{itemize}
\item How to design and develop a method to convert GPS traces into a complex 
network representation without loss of geopositional data?  
\item Are the complex network outcomings useful to provide more contextual data 
in the location of points of relevance? 

\end{itemize}
Therefore, the contribution of this paper is two-fold: on the one hand, to generate an 
information retrieval method to obtain points of relevance and the relationship between them. 
On the other hand, to apply complex networks theory to a real-world geospatial scenario.

\section*{Trail network model}

A network is made up of vertices or nodes which are connected by edges. In the modelling process, 
we need to provide semantics in each graph element since the topological features must be associated 
with real indicators of the modelled environment.  In our model, nodes are trail crossing locations, 
forks or intersections, and trailhead places. 
Trail crossings locations are critical points since they can involve  accessibility to different areas, 
change of direction, encounter with other users, or a link to other transportation system. 
Furthermore, the identification of these points are  crucial because they are places to perform interventions, 
influencing, for example, the distribution of visitors~\cite{Coombes2008}.
Trailhead places require some effort in terms of transport, time or supplementary services to reach them.  
The edges represent in our model the hiking activity connecting the nodes.

\vspace{0.2 cm}
\textbf{Definition: trajectory}\\
A trajectory ($\tau$) is a finite, time-ordered sequence of coordinates $\langle c_1,c_2,...,c_n \rangle$.

\textbf{Definition: trajectory vector}\\
A trajectory vector ($\overline{\tau}$) is the sequence of vectors between consecutive coordinates: \\$\overline{\tau}$ = $\langle v_0,v_1,...,v_{n-1} \rangle$, where $v_i=\overline{c_ic_{i+1}}$.

\textbf{Definition: intersection}\\
An intersection operation ($\bot$) between two trajectories  gives a set of ordered coordinates: $\tau_1 \bot \tau_2 = \overline{\tau_1} \bot \overline{\tau_2}  = C$ and $\forall c_i \in  C : c_i \in \{\tau_1 \cup \tau_2 \})$. \textbf{C1}: The union of consecutive points defines a common segment.  \textbf{C2}: A loop is an intersection operation in the same trajectory: $\tau_1 \bot \tau_1$
  
\begin{figure}[!h]
   \centering
  \includegraphics[width=0.65\linewidth]{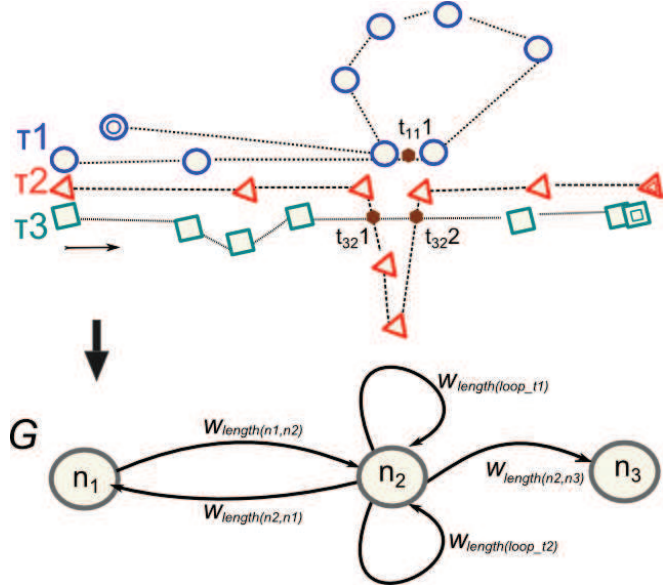}
  \caption{{\bf Sketch of the network generation from GPS traces.} Upper panel shows the track points of three GPS traces where coordinates are represented by geometric symbols (circles, triangles, and squares) and the crossing points are in brown colour. Lower panel is the directed network generated with different edge weights ($w$).}
   \label{fig1}
 \end{figure}

  
We combine all the trajectories to generate the network. 
From the perspective of a trajectory, we can extract start, end, and loop points. With the combination of several trajectories, we can detect intersections among them. 
This process is represented in Fig.~\ref{fig1} that contains three GPS trajectories  (hiker's records where the sequence  of coordinates goes from left to right) with positions biased under unknown recording conditions and the generated graph (\emph{G}). The three trajectories should be overlapped but this will difficult the interpretation process. Trajectory 1 ($\tau_1$)  is a circular route, and possesses a loop,  which represents a intersection of two segments,  a start and end points. Both points are considered as identical. The considerations on whether two points are identical in space depend on the GPS accuracy and we explain it later. Trajectory 2 ($\tau_2$) contains a loop at the same point that $\tau_1$-loop but the segment is at another location. The start and end points do not match, so it is not a circular trajectory. Trajectory 3 ($\tau_3$) is a non circular route, where the start and end points coincide with $\tau_2$ points.
The result is a directed graph: ($G$), where nodes $n_1$ and $n_3$ are produced by the start and end points of all routes. 
The $n_2$-node is determined by the origins of each loop ($\tau_1$ and $\tau_2$) and by two intersections: $\tau_1 \bot \tau_3$ and $\tau_2 \bot \tau_3$. The weight of each node is the number of traces that match in that place. 
Following the example, the weight of $n_1$ is $W_{n_1}=3$, $W_{n_2}=3$ and $W_{n_3}=2$.

Edges are defined by the sequence of nodes. 
The $n_2$-node possesses two different loops, each loop represents an alternative route. 
The weight of each edge is the number of routes between the nodes.

The graph depends on the number of samples. For example: (I)  if we only consider $\tau_3$, there is no $n_2$-node. 
(II) If we do not consider that the $\tau_2$ does not have a loop and there is no $\tau_1$, 
the intersection between $\tau_2$ and $\tau_3$ will define $n_2$-node.

 \subsection*{Algorithm}
\begin{figure}[!h]
	\centering
	\includegraphics[width=1\linewidth]{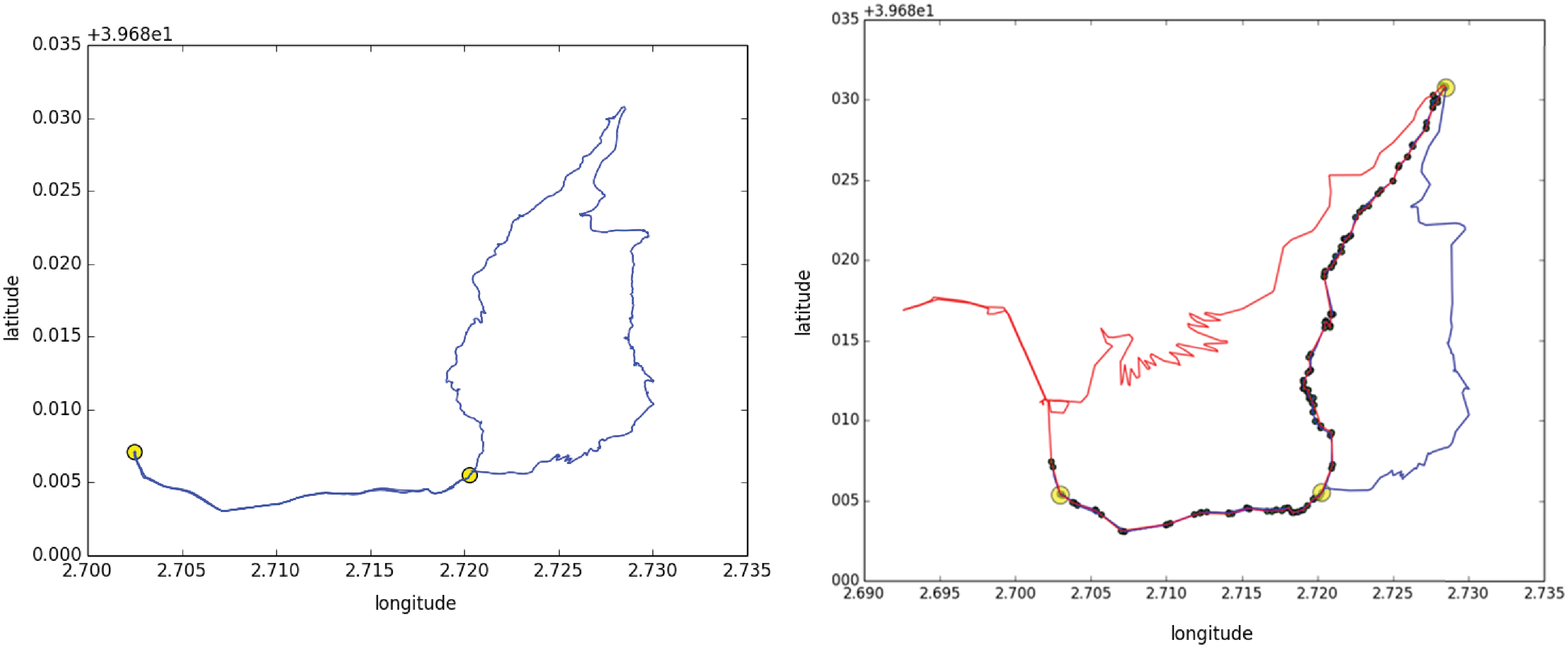} 

	   \caption{{\bf Example of the detection of candidate nodes.} (Left) The first part of the algorithm detects the start, end, and intersection points indicated by yellow circles of the trajectory.
	   	(Right) The second part of the algorithm, detects the intersection points between two trajectories. Black points define the common segments and yellow points are the candidate nodes.}

	\label{fig2}
  \end{figure}

The time complexity of a brute-force algorithm that computes the intersection 
between two pairs of coordinates using the whole dataset is $O(n!)$, where $n$ 
is the sum of coordinates of all trajectories.  
To reduce the number of cases due to GPS accuracy and several device configurations, we decide to split the process of computing 
intersections in two parts. These intersection are candidate nodes in the modelling of the network because it is necessary to consider  all of them to valid the final nodes.
 We present a pseudo-code of the algorithm in alg.~\ref{alg1}. The algorithm has two main loops. 
In the first part (lines 1-10), we individually analyse each trajectory to obtain: trajectory bounds, start and 
end points and a new smoothed trajectory ($\tau'$).  
If start and end points are approximately 
equals ($threshold_{length}$), both points are considered equals (line 5). 
We filter the trajectory using Ramer-Douglas-Peucker algorithm~\cite{Douglas73} obtaining 
a simplified trajectory $\tau'$  (line 6). After that, we compute the loop points (line 8).
In this case, the route contains an overlap 
with itself. For instance, a route that goes back the same way.  First and last 
point of each segment are estimated trail crossings, omitting nearby places 
($threshold_{sq}$).  In Fig.~\ref{fig2}(left), we show an example of a route with 
a common segment.
In this case, we detect two nodes (in yellow colour) applying the first part of the algorithm.
Left yellow point is both the beginning and the end of the 
trajectory, and at the same time, it is the start point of the common segment.  Right yellow
node is an intersection and is the other end of that segment. 
In the second part of the algorithm (lines 11-19), we only compute the intersections 
of smoothed trajectories that have an overlap of bounds. As before, we 
proceed in the same way with the sequences choosing start and end points. 
Fig.~\ref{fig2}(right) shows an example of this second part of the algorithm. 
In the overlapping of both trajectories, there are some common points which are in black colour. 
The union of these points defines various segments using a threshold of separation ($threshold_{sq}$).
In this case, selected nodes are the endpoints of the sequences (in yellow colour).

 \begin{algorithm}[!h]
   \caption{Computing Candidate Nodes}\label{alg1}
   \begin{algorithmic}[1]
   \Require trajectory dataset $T$, $Threshold_{length}$, $Threshold_{sq}$
   \Ensure intersection points $IS,IB$; start/end points $sC,eC$
   \For{ each $\tau_i\in T$ }
      \State  $ B_i  \gets$ get Bounds  ($\tau_i$)
      \State $ sC(\tau_i.id,i)  \gets$ get First Coordinate ($\tau_i$)
      \State $ eC(\tau_i.id,i)  \gets$ get Last Coordinate ($\tau_i$)
      \State \textbf{If} distance($sC_i,eC_i)<=Threshold_{length}$\textbf{:} $sC(\tau_i.id,i)  \gets eC(\tau_i.id,i) $
      \State $\tau_i' \gets$ do a Smooth Filter  ($\tau_i$) 
      \State $T' \gets$ insert($\{B_i,\tau_i'\}$)
      \State  $ ipoints_{i}  \gets$ ($\tau_i' \bot \tau_i'$)
      \State $IS(\tau_i.id,i) \gets$ compute Start/End points of each sequence $(ipoints_{i}, Threshold_{sq})$
   \EndFor
   \Repeat
     \State $\{B_i,\tau_i'\}  \gets pop(T')$
      \For{ $\{B_j,\tau_j'\}  \in T'$ }
        		\If{$B_i \cap B_j \neq\emptyset$}
        			 \State  $ ipoints_{ij}  \gets$ ($\tau_i' \bot \tau_j'$)
        		\EndIf
        		\State $IB(\tau_i'.id,i,j)) \gets$ compute Start/End points of each sequence $(ipoints_{ij}, Threshold_{sq})$
     \EndFor
   \Until{$T'.size == 0$}
   \end{algorithmic}
 \end{algorithm}

The intersection process (lines 8 \& 15) uses a \textit{k}-d tree algorithm to 
facilitate the nearest point matches~\cite{Friedman1977}.  The idea is to detect the longest common subsequence between two trajectories~\cite{VlachosGK02}.   In this way, 
intersection points are obtained from real coordinates, they are not obtained from an 
interpolation process or an average computation. The algorithm requires a 
threshold to compute the closeness and returns a ordered sequence of 
coordinates. In the second call (line 15), the algorithm avoid nearest points of 
the same route since they are obtained in the first loop. We make publicly available this part of the  algorithm~\cite{refAlgorithm}.

\begin{figure}[!h]
    \centering
    \includegraphics[width=.5\linewidth]{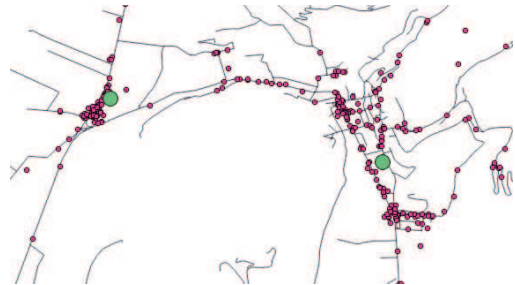}
    \vspace{0.3 cm}
    \caption{{\bf Illustration of the obtained centroids.}  The start, end, and fork points in Bunyola (Majorca) 
    are represented in purple. After the mean shift algorithm, two centroids are obtained (green points). 
    Black lines correspond to roads.}
    \label{fig3}
  \end{figure}

Depending on the distribution of coordinates among trajectories and on the 
selection of thresholds, there may be a random distribution of points around a 
real trail crossing.  Thus, we need to group nearby 
points when all intersection points of the whole set of 
trajectories is calculated in previous algorithm. We compute centroid points applying a mean shift clustering 
algorithm~\cite{Comaniciu02meanshift}. The centroids are the nodes of the 
network, and theirs weights are the number of GPS traces that goes there. Returning 
to the example of Fig.~\ref{fig1},  if we assume that 
selected thresholds are exceeded due to distance issues then the intersection 
process ($\tau_2 \bot \tau_3$) provides two intersection points: $\tau_{32}1$ 
and $\tau_{32}2$. At the same time, $\tau_1$ has a intersection point 
($\tau_{11}1$) for its loop which is physically collocated between both. 
Thus, we have three possible points to represent an unique trail crossing: 
$\tau_{32}1$, $\tau_{11}1$ and $\tau_{32}2$. In Fig.~\ref{fig3} we 
show an example of the dispersion of  candidate nodes. 
After the application of mean shift algorithm, we obtain the centroids of the cloud of candidate nodes. These centroids are the nodes of the generated network.

The edges are the sequences of candidate nodes of each route.  Before the 
clustering process, we have a sorted sequence of nodes of each route: start 
point, n-intersections (loops and intersections among trajectories) and end 
point. The range of nodes of each trajectory is (1..$n$). We transform this 
sorted sequence in edges by mapping the points in the correspondent final 
centroids. We discard edges with the same source and target node since they 
represent nearby intersections, and sequence length values equal to one. The 
weight of each edge is the length between nodes considering the original 
trajectory.

\section*{Case study:  hikes in Balearic Islands}

Our dataset consists on GPS traces from a sport social tracker application that 
is used as a collection of routes~\cite{refWikiloc}. The application only 
stores the same route once per user.  The recorded data represents the real human 
movement in any localization (urban, interurban and natural areas), and reflects 
external and personal factors in every individual and collective records. 
Outdoor activities are conditioned by external factors such as weather 
conditions, daylight hours, holidays, specific planning trails, social events, 
etc. These meaningful data are registered by different devices and 
configurations which affect the distribution of points and their accuracy. 
For each GPS trace, we have the  following metadata: start and end timestamps, the degree of difficulty of the route  assign by the user, and a GPX file with the coordinates of the track points.
 We make publicly available the anonymized set of GPS traces~\cite{refDataSet}.  In total there 
are 15376 records (21.2 millions of coordinates) performed by 2965 users over 8 years 
(from 2009 to 2016) on the Balearic Islands (Spain).  This archipelago is interesting for the study by the following reasons: (I) they are relatively small confined pieces of land where the spatial dimensions are closed to the human movement scales, (II)  the three main islands possess very different geographical characteristics, (III) the weather conditions allow for the practice of sport activity during all the year,  and (IV) a future applicability of findings in well-known tourist region in Europe. A more detailed description of the GPS traces can be found in the Data Description section in \nameref{S1_File}.


\subsection*{Basic statistics}
We introduce the following statistical measures about weather conditions and 
topography of our scenario in order to clarify the results of the network 
analysis.

\subsubsection*{Weather conditions}
\begin{figure}[!h]
   \centering
    \includegraphics[width=1.\linewidth]{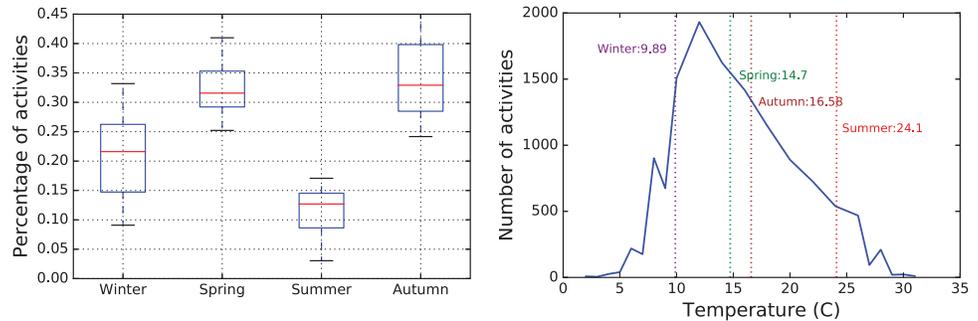}
      \caption{{\bf Analysis of hiking activities. } Seasonal frequency of hikings for the four seasons (left). 
      The number of hiking activities 
      as a function of the average temperature of the day (right). Vertical dotted lines represent the averaged seasonal temperature.}
   \label{fig4}
   \end{figure}

Outdoor activities are influenced by weather conditions. Balearic Islands have 
a Mediterranean climate: warm winters, and hot and humid summers.  The 
average temperature in August is 25.9$^{\circ}$C  with an 
average maximum temperature of 29.5$^{\circ}$C. In contrast, 
in  January the average temperature is 11.7$^{\circ}$C  with an 
average minimum of 8.3$^{\circ}$C. We obtained the temperature log from~\cite{wunderground}.
This range of temperature has a direct consequence in the number of activities as 
Fig.~\ref{fig4}(left) shows. In summer, the number of activities is 
significantly lower than in other seasons, due to high 
temperatures. This is clearly seen in Fig.~\ref{fig4}(right) in which the 
number of activities decreases when temperatures increase.  

\begin{figure}[!h]
  \centering
   \includegraphics[width=1.\linewidth]{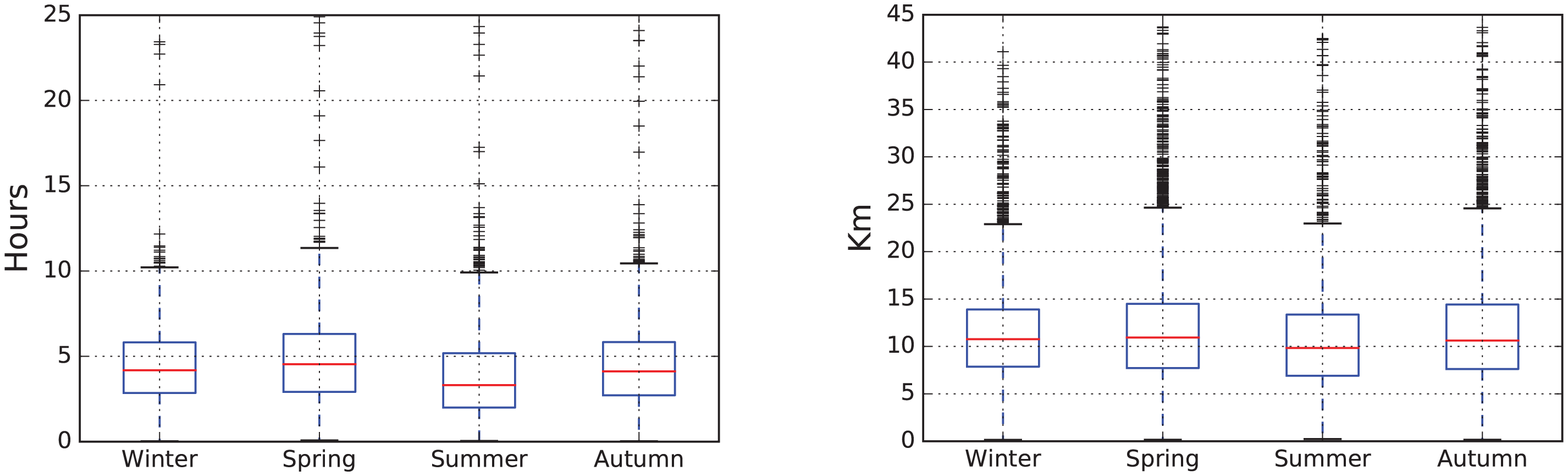}
    \caption{{\bf Seasonal analysis of hiking duration and length.} Box plot figures: duration (left) and length (right).}
  \label{fig5}
  \end{figure}

The number of daylight hours also influences in the behaviour of the hikers. In summer the 
number of daylight hours is approximately 14 hours, however, in winter it is around 
10 hours. Fig.~\ref{fig5} (left) shows the distribution of the duration of 
activities for each season.  We can observe that the temperature has a greater 
influence than daylight hours in order to determine the duration of activities. 
Furthermore, there are many routes with a mean duration below 5 hours; and in 
spring, there are a small group of striking routes lasting more than a day, 
which correspond to long-distance footpaths. The length of the hikes is 
represented in Fig.~\ref{fig5} (right).
   
 \subsubsection*{Topography}
 
\begin{figure}[!h]
 \centering
 \includegraphics[width=1.\linewidth]{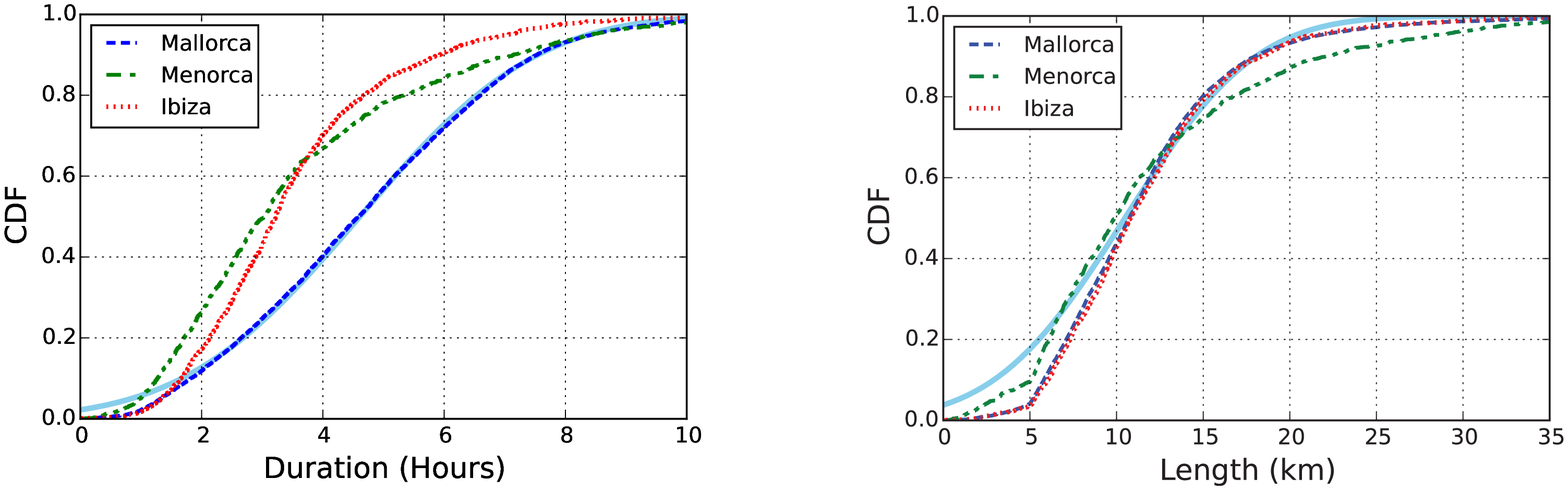}
\caption{{\bf Hiking duration and lenght.} Cumulative distribution function of the hiking duration (left) and 
hiking length (right) in each island. In both panels, solid line represents a fit to a cumulative Gaussian distribution: 
$f(x)=1/2[1+erf((x-\mu)/\sigma\sqrt 2)]$ with mean $\mu=4.6 \;(10.87)$ and standard deviation $\sigma=2.3 \;(4.57)$ 
for the duration (length) distribution in Majorca.}
\label{fig6}
\end{figure}

The topography of each island influences also in the hiker activity . We  
consider the four main islands: Majorca, Minorca, and Ibiza \& Formentera.  
Majorca has a  lower-intermediate mountain range called \emph{Serra de 
Tramuntana} with several peaks over 1000 meters where Puig Major is the highest 
peak (1445 m), the most important trail is GR221 (\textit{Grande Randonn\'ee}) 
with a length of 135 km. Minorca has a well-known trail called 
\emph{Cam\' i de Cavalls} or GR223 which is a circular trail with a length of 185 km 
around the perimeter of the island. Finally, Ibiza trails are 
scattered among the small villages.  We include Formentera island in the same analysis than Ibiza. From now on, we only mention the three groups, or the three islands.
Fig.~\ref{fig6} shows the cumulative distributions of the duration and length of the hiking activity in each island. 
The distributions of the length are very alike indicating similar hiking length in the three islands. 
However, the distributions of durations exhibit substantial differences, significantly in Majorca showing a wider distribution. 
Although the  hiking length is similar in the three islands, in Majorca the hikes tend to have a longer duration. 
The distribution of duration (length) follows a Gaussian distribution with mean $\mu=4.6 \;(10.87)$ and 
standard deviation $\sigma=2.3 \;(4.56)$ for Majorca.

\section*{Analysis of the hike network}

\subsection*{Trail network}

To analyse the Balearic Islands trail network, we configure the previous algorithm with the following parameters: $threshold_{sq}=3$, $threshold_{length}=10$ meters, \textit{k}-d tree query distance = 55 meters  and a mean shift bandwidth of 0.0043 (approx. 480 m).  These values are defined through empirical analyses on different regions of the islands. The value with more sensitivity in the features of the network is the bandwidth of the mean shift algorithm. Setting this value is a trade off among the size of the region, the physical characteristics of the area and the level of resolution of trail crossings. We set to have a resolution of a half a kilometre. In any case, the topological features remain proportional up to a value above two kilometres.  The k-d tree query distance allows the differentiation of nearby paths. Paths less than 55 meters are considered similar.  Finally, we choose a sequence of three sampling $threshold_{sq}$ in the overlapping of the paths. This value is close to 60 meters, enough  to differentiate nearby paths.


We observe that the number of candidate nodes is high with a huge volume of 
traces. There is an effect of sequencing intersection points due to the 
dispersion of coordinates. The problem of finding the intersection of 
trajectories has become a problem of finding the real trail crossing. The number 
of intersections of a route is so high that this number exceeds the number of 
coordinates of a route. The average ratio of intersection points 
is approximately twice the number of coordinates on the route. For instance,  the 
trajectory shown in Fig.~\ref{fig2} (left) has 1054 coordinates for a length 
of 12.3 km, the number of intersections points with the rest of 
routes (approx. 1.4k in this area) is 3428 points. 
In order to reduce the number of possible candidates, we introduce an extra step to the previous algorithm, by 
applying  the mean shift algorithm to the node candidate of each traces. Then, the same algorithm is applied on all 
candidate nodes of all traces. Now, the position of the nodes gives us an 
approximation for the real intersection. 
 
\subsection*{Overview}

We characterize the network resulting from the algorithm previously described by 
computing the standard properties summarized in Table~\ref{table:summary}. These 
values are obtained using NetworkX library~\cite{hagberg2008}. Furthermore, 
with the aim to study the seasonal variation, we consider three different 
segmentation of the data analysing the following seasonal networks: spring (SP), 
summer (SU), and the aggregation of the four seasons (4S). 
These networks have different number of isolated subgraphs,  which value varies seasonaly on each island. 
The number of subgraphs is an indicator of the distribution and use of places. 
The main subgraph of Majorca is localized in the Serra de Tramuntana area,
where there are sections of the GR221. In Majorca and Ibiza, we 
observe greater variability of subgraphs in each season.

The number of activities decreases in summer in all areas, but this trend is lower 
in coastal areas. In any case, Minorca is the only example that maintains 
approximately the same localizations of nodes in each season. In addition, 
Minorca presents special circumstances: some of the coast lines and beaches are 
only accessible by going bordering private lands.   

\begin{figure*}[!h]
  \centering
  \includegraphics[width=1.48\linewidth]{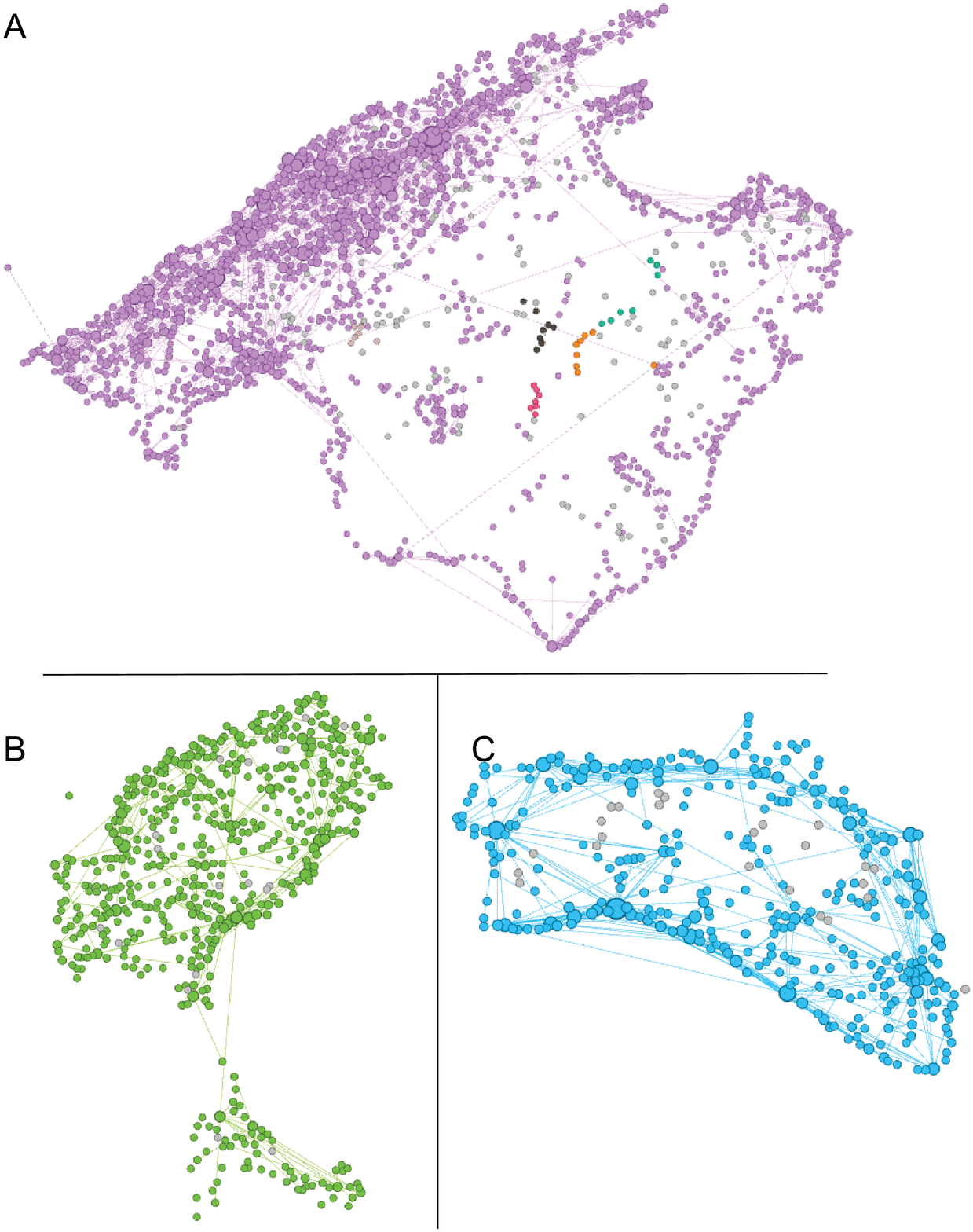}
  \caption{{\bf Hiking activity networks in the Balearic Islands obtained from algorithm~\ref{alg1}.} 
  Majorca (A), Ibiza \& Formentera (B), and Minorca (C) networks correspond to the aggregation of the four seasons. 
  Black nodes represents small subgraphs (number of nodes $<4$).  In addition, the size of the nodes is proportional to 
  the number of different hikes in that area.}
	   	   
\label{fig7}
\end{figure*}

The extracted networks for Majorca, Minorca, and Ibiza \& Formentera aggregated for the four seasons are shown in Fig.~\ref{fig7}, respectively.  In Majorca island, the nodes are concentrated in the mountain area. In Minorca, the nodes are around the perimeter. In Ibiza \& Formentera, the localization is more uniform. 
The locations of nodes change according to the season, except the nodes with more weight which remain relatively in the same area. We experience that the high 
temperatures in summer make coastal areas more attractive. 
We include  the spring and summer figures of the networks of each island in the section of Hiking activity networks in \nameref{S1_File}. In addition, networks can be found in the Supporting Information (\nameref{S2_File}). 


\begin{landscape}
\begin{table}[!ht]
\centering
\caption{
{\bf Summary of topological features of the three networks.}}
\begin{tabular}{l| l l l  | l  l l | l  l  l }
  \textbf{Property} 		&  \multicolumn{9}{c}{} \\
   \multicolumn{1}{r|}{Island} 	&  \multicolumn{3}{c|}{\textbf{Majorca}}&		 		\multicolumn{3}{c|}{\textbf{Minorca}}&   					\multicolumn{3}{c}{\textbf{Ibiza \& Formentera}} \\
    \multicolumn{1}{r|}{Area} 	&  \multicolumn{3}{c|}{3640 $km^2$}&		 		\multicolumn{3}{c|}{702 $km^2$}&   					\multicolumn{3}{c}{654 $km^2$} \\
  \multicolumn{1}{r|}{\textbf{Season} } 			& \textbf{4S} &\textbf{ SP} & \textbf{SU}	   & \textbf{4S} & \textbf{SP} & \textbf{SU}		 & \textbf{4S} 			& \textbf{SP} & \textbf{SU}\\ 
    \multicolumn{1}{r|}{Number of traces} 	&  11984	&	3964	   &  1209				   &  1492    	&  427       & 390         				   			     &  2243		&   702       			 &   277     \\\hline 
  Number of nodes 	($n$)     & 	 884   &	784		&   767				  & 203    &    161     &   218  						  		 	 	& 293		&         246      		&   210      \\ 
  Number of edges ($e_m$) 	     & 	 10864 	  &	3119	&   1225					  & 1616   &  314      &  475   						  		 	 &	1944		&     454          		&   146     \\
  Degree ($k$)      		  & 14.943 	&	7.3083	   &  4.4067  				&   11.3899       &   4.9507           &  6.0041         	&	8.0469	&         4.2651  			 &  3.2426    \\ 
  Degree range      		  & (1,22) 	&	(1,22)	   &  (1,22)  				&   (1,24)       &   (1,7)          &  (1,24)         				   	  &	(1,28)		&         (1,24) 			 &  (1,7)    \\ 
  Avg. shortest path       &  5	 &	5	     &   	8									 & 3         &  6           &    4    							 		  &		4	 &     4                  &   4     \\
  Diameter 	                    & 16	&	18	   &    25							       &  9        &  11         &  11         				   			      &	 10		  &    18       			 &   11     \\ 
 
  Avg. clustering ($C$)       & 0.342443	& 0.313013  &  0.199265   & 0.321705  & 0.137489    & 0.207485						&	0.3152	& 0.189795   & 0.198618\\
  Avg. weighted clustering $(C_i^W)$ & 0.011435    &	0.008306 &   0.005601 	& 0.030659 & 0.011814     &  0.015959&		0.0221	 			 & 0.011479   &  0.016277 \\
  Assortativity ($r$)            & 0.138867    &  0.121049 &  0.022  		 & 0.027258  &  -0.051515 &   -0.0127    				&	0.0044      & -0.142534 &  0.273333\\
  
  Avg. edge weight (km)       & 5.47	      &	4.023		&   6.43					& 7.03    & 6.44   &   5.12  						 			   &	4.02	    & 4.16       			 &  6.44      \\
  Edge weight range (km)      & (0.39,86.6)	  &	 (0.39,65.3) &  (0.39,65.2)    &  (0.3,42.2)     &  (0.37,18.9) &  (0.4,37.3)             &	(0.6,34.38)		&   (0.4,37.2)      & (0.37,18.4)        \\
   
  N. of subgraphs ($n>3$)  & 5	&	5	   &    3							        &  1    	&  1           &  1         				   			      	&	1	&    1       			 &   4     \\ 
  N. of isolated nodes 	  & 	73   &	 156		&   	223					   & 15       &  26         &     30   					  		 	 		 &   15    &          41     		&    127    \\
  N. of unique edges ($e$)	 & 	 2567 	  &	1314		&   764				 & 595     &   202      &  307   						  		 	 &	706		&     291          		&   96     \\
  
\end{tabular}
\begin{flushleft} Results of the three islands considering all seasons (\emph{4S}), and both most extreme seasons in terms of frequency of activities : spring (\emph{SP}) and summer (\emph{SU}).
\end{flushleft}
\label{table:summary}
\end{table}
\end{landscape}

\subsection*{Basic network properties: nodes and edges}

As we mentioned, nodes represent relevant areas where intervention tasks could be executed. 
The node weights are the number of activities that have been 
performed in that area. From table~\ref{table:summary}, we observe that the number of 
nodes is slightly similar during spring and summer within each island. The 
difference of nodes between seasons does not depend on the number of samples. 
Thus, we can infer that the number of visited areas does not change many among seasons 
despite of the reduction in the activity frequency and the change of the localizations. 

The locations of  nodes with bigger weights remain the same among seasons while
the locations of nodes with lower weight change from inland to 
coastal areas in summer. It is remarkable the absence of nodes in the inland area of 
Majorca during the summer.  Moreover, we highlight the importance of GR trails 
since most weighted nodes are localized there. In Ibiza, the non-existence of 
guided hikes may lead to greater dispersion of trailheads and forks.

The edges are an indicator of the connectivity of segments of multiple trails 
that a node possesses although the number of edges depends on the intersections 
of a route. If a trajectory does not intersect with another or does not have 
loops, the resulting graph will only have the start and end nodes. Thus, a set of 
connected nodes represents the utilization and existence of a network of paths 
linking all these locations.  In Table~\ref{table:summary}, the \emph{number 
of edges} ($e_m$) represents the edges in the whole network 
and the \emph{number of unique edges} ($e$) is the value using a 
directed graph.  We know that the average number of loop hikes is around  59.9\%, then,
we can estimate the average number of intersections of one trajectory with 
another. This indicator is the ratio between the number of unique edges $e$ and 
nodes ($n$) : $e/(n*59.9\%+n*40.1\%*2)$. This value is 2.073 in Majorca, 2.09 in 
Minorca, and 0.71 in Ibiza \& Formentera. 

The existence of disconnected nodes generates isolated subgraphs. The number of 
isolated subgraphs increases in summer season due to the lower number of 
activities and a reduction in long distance routes.  Long routes lead to a 
greater probability of intersection with others.  This indicates that the 
temperature influences the attractiveness of the location and the utilization of 
places.

\subsection*{Degree and connections} 

The degree of a node is an indicator of the connectivity 
among different areas.  The average  degree is different in each island and it decreases 
from sprint to summer in Majorca and Ibiza. In contrast, in Minorca, the 
average degree increases during the same period. In general, the connectivity 
decreases in summer except in Minorca where the presence of nodes is still 
intensified in the coastal areas.  

\begin{figure}[!h]
  \centering
  \includegraphics[width=.7\linewidth]{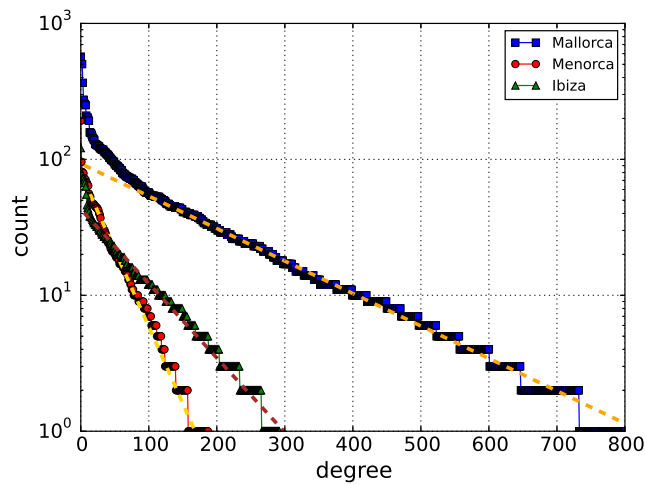}
  \caption{ {\bf Node degree distribution.} The degree distribution for the three islands for the aggregated 4S. 
  Dashed lines represent fits to  $f(x)\sim exp(-bx)$ with $b=(0.005, 0.026, 0.012)$ for Majorca, Minorca, 
  and Ibiza \& Formentera respectively.}
  \label{fig8}
\end{figure}

Fig.~\ref{fig8} shows the degree distribution for each of the three islands for the 4 season aggregated network. Majorca shows a larger degree values while Minorca and Ibiza present similar values. This indicator shows the same evolution in other seasons among the three islands. The degree distribution of the three islands follows an exponential distribution. Similar behaviour has been observed in different transport networks, for example, in the Singapore's bus network~\cite{Soh2010}.

The weight of an edge represents the straight distance between two nodes. Thus, 
the real segment of a trail that joins two nodes can possess bigger length than 
the weight of the edge. The weight value grows in the summer season in Majorca and Ibiza. 
The lower the density of trajectories, the smaller number of intersections, and consequently, the shorter the lengths between 
them. The range of weights is an indicator of the minimum and maximum lengths of 
the trails in each island. In the case of Majorca, the maximum distance 
corresponds to the \emph{GR} trail length.

On the other hand, we compare the possible relationship between two indicators 
conceptually close: the weight and degree of a node.  We use Pearson correlation 
coefficient to compare both. In spring, the correlation values are 
the following:  0.9374 in Majorca, 0.4157 in Ibiza, and 0.9612 in Minorca. In summer, the values are: 0.8281 in Majorca,  
0.1221 in Ibiza, and 0.9612 in Minorca. 
This implies that both features are an indicator of the number of 
alternative hikes of a given place: the larger the degree of choices, the 
greater the number of alternative recorded trails. We are able to detect places 
with a high number of alternatives using topological features. 
We observe that the position of 
the main nodes in terms of degree, weight, and their locations matches with 
well-known places e.g: water reservoirs (Cuber and Gorg Blau), mountains towns (Escorca, Esporles), to mention some of them.

The average shortest path is also affected by the attractiveness of coastal 
areas in summer. In Majorca, the average shortest path is 5 in spring, and 
it increases to 8 in summer. In Ibiza is identical in both seasons. However, in 
Minorca the value decreases in summer since there are more  
activities around the coastal areas and then, the nodes are better linked. The 
diameter is defined by the maximum shortest paths between two nodes. In terms of 
hiking, it is representative of the longest route that hikers can make and it 
depends on the localization of the nodes.  Majorca has a larger diameter than 
the rest of the islands. In Minorca, the diameter remains equal between both seasons 
since the locations of nodes remain around the perimeter of the island. 
In Ibiza the diameter decreases from spring to summer due to the increase in the number 
of isolated subgraphs.

\subsection*{Community analysis}

\begin{figure}[!h]
 \centering
 \includegraphics[width=1.\linewidth]{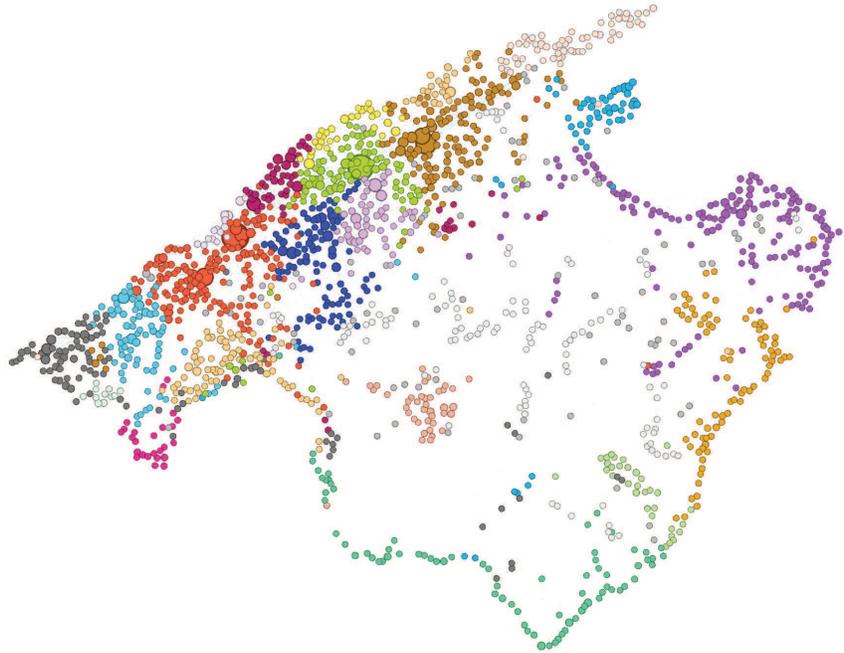}
\caption{{\bf Community analysis of Majorca Island.} Different colors indicates the largest communities. 
Communities with size smaller than 20 nodes are colored with light gray.}
\label{fig9}
\end{figure}

We perform a community analysis to the 4S network of Majorca island using the Louvain method~\cite{LouvainMethod} 
implemented in Gephi ~\cite{ICWSM09154}. Fig~\ref{fig9} shows colored the largest communities found. 
The communities segment the island by in  areas revealing the distribution of hiking activities. 
The community analysis also reveals the area of influence of the main points of relevance. 
For Minorca and Ibiza \& Formentera a big dominant community is found in each island. 

\subsection*{Clustering and correlations} 
Clustering coefficient ($C$) is a measure of cohesiveness around a node. It takes values from [0,1] where 0 and 1 indicate that none or all neighbouring 
nodes are linked~\cite{Szabo2004}. The highest clustering values correspond to nodes with the highest number of hikers.
In Table~\ref{table:summary}, the average clustering 
coefficient has a range from 0.13 to 0.31. We contextualize this value comparing 
it with other network models.  For example, Singapore rail system possesses a value of 
$C_{Sing.}=0.934$~\cite{Soh2010}. Poland transport systems has a range of  
$0.68 < C_{Poland} < 0.85$~\cite{Sienkiewicz05}. Our case of study is 
slightly above to the Portland network $C_{PL}$) = 0.0584~\cite{Chowell03}. It 
is desirable that a transport network is designed to minimize the number of 
exchanges and maximize the use. The network of trails were built to connect 
areas or to achieve natural resources (i.e. cultivated fields, hunting trails, 
etc.).  In our study, we observe that the clustering cohesiveness is formed by edges with low weights.

Another topological property of the network is the type of correlation among 
nodes. A network is assortative whether  high-degree nodes have a trend  to 
connect to other high-degree nodes~\cite{Newman}. Otherwise, a network is disassortative  
whether the  low-degree nodes tend to connect to high-degree nodes. Assortativity 
takes values in the interval [-1,1] where -1 indicates a dissortative 
network and 1 indicates an assortative one. The assortativity of the studied networks depends on the season 
especially in the two small islands. In Majorca, the assortativity tends to 0 in 
summer. 


\section*{Discussion}

The novelty of this study is the applicability of complex network theory to 
trail networks. The exploration of trail networks through hiker activities offers 
a real vision in the understanding the use of natural resources.  In our case,  the agglomeration of intersections of  trajectories and trailheads define the network.  This 
theory is already applied in transportation systems such as Singapore RTS and 
BUS~\cite{Soh2010}, Boston subway~\cite{Latora2002109}, Chinese 
railway~\cite{Li2007}, world-wide airports network~\cite{Amaral2000} and 
Beijing traffic road network~\cite{TIAN2016537}. In most of the cases, the 
interpretation of the network is defined by the two basic matching rules: the  
``transport stop places'' are nodes and the connections  or movements of 
passengers among the nodes are edges. The weight of each edge can have different 
values: the number of passengers, the distance, the duration of the trip, etc.

A typical trajectory obtained from GPS devices is a sequence of 
longitude, latitude, and timestamps. Modern devices include 
more information: velocity, heart rate, elevation, slope, etc.  Unfortunately, 
the quality relies on the precision of the coordinate acquisition from satellite 
signals. The error and the size of the sequence are relevant to match them to 
physical areas and, for instance, to decision making regarding a 
change of direction between other analysis. In the case of mobile phone connections, the sequence of the 
trajectory is set by the carrier's antennas. A GPS trajectory can possesses more 
intermediate points but they are subject to external constraints (e.g. natural 
or urban canyons, atmospheric conditions, time to acquire fix, etc.) or to user 
forgetfulness (e.g. battery charge, keeping in bag, auto travel before acquiring 
fix, etc.)~\cite{Moreira2010}. Filtering and other statistical methods can be 
applied to mitigate the error. Each trajectory can be processed to reduce noise 
avoiding outlier coordinates or interpolating points. However, the position and 
velocity  are typically biased and have unknown distributions. If the goal of 
the study is to discover specific areas, the homogenization of points may not be 
effective. Other group of techniques called map-matching establishes relationships 
among the coordinates with real information coming from maps. In this way, it is 
determined which road segments have been used in each trajectory. It is not an 
easy task since it depends on the velocity of the vehicle, the frequency of 
samples, the precision of the position and the density of 
roads~\cite{Brakatsoulas2005,Shang2014}. We address our approach to the 
identification of trail crossings without map information. Collaborative 
resources such as Open Street Map provides trail networks but our goal is only 
to depend on the network that is traced by hikers. For a future work, we want to 
compare our model with the  network generated from this spatial sources.


\section*{Conclusions and future work}

In this article, we present a novel approach to exploit records of sports 
activities through the generation of a hiking activity network and the application of 
complex network theory for identifying points of relevance. 
The obtained points of relevance, together with the extracted topological features,
reflect the utilization of trail segments and trailheads where environmental 
management plans can be carried out.  
In addition, our method relates places among them in contrast with other approaches where 
the regions or points of interest are obtained as isolated elements.  
In the modelling of a network, it is 
important to consider the semantics of its elements, therefore we decided to 
model nodes as trail crossings. The use of trail crossings is a powerful concept 
to describe passageways and redirect the exploitation of natural spaces. In the first part of this article, we 
propose an algorithm to detect trail crossings from GPS trajectories which 
generates the network.  As the generated network depends on the 
activity of the users, this study is also the first step of a global analysis 
of human outdoor activities considering geographical and topological aspects.

We detect that weather conditions affect outdoor activities 
and we reflect this fact in the construction of different networks according to the season of the year. 
Thus, we generate three networks considering the aggregation of trajectories in the four seasons, summer and spring seasons, 
which are the most extreme  in terms of frequency of activities,
analysing for each network the main topological features.
From the localizations of nodes, we observe for the three islands that the number of visited areas remains similar during the 
year but there is a reduction in the number of frequencies and  a relocation 
from inland regions to coastal areas in summer season. Regardless the season, 
the segments of the long trails (\textit{Grande Randonn\' ee}) have a high density of 
nodes and do not lose importance in topological terms. For summer in the case of Minorca island, 
the nodes tend to cluster together along the perimeter of the island 
where the GR is located.  
The community analysis performed in Majorca island allows the identification of 
the area of influence of each point of relevance. 
From the degree feature, we detect that the attractiveness of a place 
depends on the number of connections, that is, the number of alternative routes. 
To sum up, the interpretation of each feature provides indicators on the utilization of regions and can be used in forecasting models.

One limitation of this study is to ascertain with high precision the discovery of trail crossings among all the 
samples. Our resolution is a local 
approach considering sequence of  pairs of coordinates with the same route and 
pair of routes. As future work, a global approach should be considered the whole 
dataset; from this perspective it can rule out  slight variations of the course 
and  inaccurate samples of GPS devices.
This study can be extended in  two directions. 
The first one is to analyse the flow of outdoor activities and their distribution throughout 
the year. 
The second one, given that this study provides measures on a precise area, is 
to analyse conflicts among other type of activities, the discovering of new paths, or 
identification of the access to restricted areas.

\section*{Supporting Information}

\paragraph*{S1 File}
\label{S1_File}
{\bf Detailed information of raw data, and hiking activity networks.} 

\paragraph*{S2 File}
\label{S2_File}
{\bf Four seasons, spring, and summer networks in GEXF format.} 

\section*{Acknowledgements}
All authors acknowledge financial support from the Departament de Matem\`atiques i Inform\`atica de la Universitat de les Illes Balears. VME acknowledges financial support through project SPASIMM (FIS2016-80067-P AEI/FEDER, UE). TP acknowledges support from the program Juan de la Cierva of the Spanish Ministry of Economy and Competitiveness. The funders had no role in study design, data collection and analysis, decision to publish, or preparation of the manuscript.


\bibliography{ref_ilc}

\begin{thebibliography}{10}

\bibitem{Song2010}
Song C, Qu Z, Blumm N, Barabási AL.
\newblock Limits of Predictability in Human Mobility.
\newblock Science. 2010;327(5968):1018--1021.
\newblock doi:{10.1126/science.1177170}.

\bibitem{cortes2014}
Cort{\'e}s R, Bonnaire X, Marin O, Sens P.
\newblock {Sport Trackers and Big Data: Studying user traces to identify
  opportunities and challenges}.
\newblock {INRIA Paris}; 2014. RR-8636.

\bibitem{Orellana2012a}
Orellana D, Bregt AK, Ligtenberg A, Wachowicz M.
\newblock Exploring visitor movement patterns in natural recreational areas.
\newblock Tourism Management. 2012;33(3):672 -- 682.

\bibitem{Beeco2014a}
Beeco JA, Hallo JC, Brownlee MTJ.
\newblock \{GPS\} Visitor Tracking and Recreation Suitability Mapping: Tools
  for understanding and managing visitor use.
\newblock Landscape and Urban Planning. 2014;127:136 -- 145.
\newblock doi:{http://doi.org/10.1016/j.landurbplan.2014.04.002}.

\bibitem{Meijles2014a}
Meijles EW, de~Bakker M, Groote PD, Barske R.
\newblock Analysing hiker movement patterns using \{GPS\} data: Implications
  for park management.
\newblock Computers, Environment and Urban Systems. 2014;47:44 -- 57.
\newblock doi:{http://doi.org/10.1016/j.compenvurbsys.2013.07.005}.

\bibitem{Beeco2013a}
Beeco JA, Hallo JC, English WR, Giumetti GW.
\newblock The importance of spatial nested data in understanding the
  relationship between visitor use and landscape impacts.
\newblock Applied Geography. 2013;45:147 -- 157.
\newblock doi:{http://doi.org/10.1016/j.apgeog.2013.09.001}.

\bibitem{Paldino2015a}
Paldino S, Bojic I, Sobolevsky S, Ratti C, Gonz{\'a}lez MC.
\newblock Urban magnetism through the lens of geo-tagged photography.
\newblock EPJ Data Science. 2015;4(1):5.
\newblock doi:{10.1140/epjds/s13688-015-0043-3}.

\bibitem{Yuan2012a}
Yuan J, Zheng Y, Xie X.
\newblock Discovering Regions of Different Functions in a City Using Human
  Mobility and POIs.
\newblock In: Proceedings of the 18th ACM SIGKDD International Conference on
  Knowledge Discovery and Data Mining. KDD '12. New York, NY, USA: ACM; 2012.
  p. 186--194.
\newblock Available from: \url{http://doi.acm.org/10.1145/2339530.2339561}.

\bibitem{Huang2009a}
Huang YT, Chen YC, Huang JH, Chen LJ, Huang P.
\newblock YushanNet: A Delay-Tolerant Wireless Sensor Network for Hiker
  Tracking in Yushan National Park.
\newblock In: 2009 Tenth International Conference on Mobile Data Management:
  Systems, Services and Middleware; 2009. p. 379--380.

\bibitem{Mulligann2011a}
M{\"u}lligann C, Janowicz K, Ye M, Lee WC.
\newblock In: Egenhofer M, Giudice N, Moratz R, Worboys M, editors. Analyzing
  the Spatial-Semantic Interaction of Points of Interest in Volunteered
  Geographic Information. Berlin, Heidelberg: Springer Berlin Heidelberg; 2011.
  p. 350--370.
\newblock Available from: \url{http://dx.doi.org/10.1007/978-3-642-23196-4_19}.

\bibitem{Giannotti2011}
Giannotti F, Nanni M, Pedreschi D, Pinelli F, Renso C, Rinzivillo S, et~al.
\newblock Unveiling the Complexity of Human Mobility by Querying and Mining
  Massive Trajectory Data.
\newblock The VLDB Journal. 2011;20(5):695--719.
\newblock doi:{10.1007/s00778-011-0244-8}.

\bibitem{Biljecki2013a}
Biljecki F, Ledoux H, van Oosterom P.
\newblock Transportation Mode-based Segmentation and Classification of Movement
  Trajectories.
\newblock Int J Geogr Inf Sci. 2013;27(2):385--407.
\newblock doi:{10.1080/13658816.2012.692791}.

\bibitem{Zheng2012}
Zheng YT, Zha ZJ, Chua TS.
\newblock Mining Travel Patterns from Geotagged Photos.
\newblock ACM Trans Intell Syst Technol. 2012;3(3):56:1--56:18.
\newblock doi:{10.1145/2168752.2168770}.

\bibitem{Brockmann2006a}
Brockmann D, Hufnagel L, Geisel T.
\newblock The scaling laws of human travel.
\newblock Nature. 2006;439(7075):462--465.

\bibitem{Hoogendoorn2005}
Hoogendoorn SP, Bovy PHL.
\newblock Pedestrian Travel Behavior Modeling.
\newblock Networks and Spatial Economics. 2005;5(2):193--216.
\newblock doi:{10.1007/s11067-005-2629-y}.

\bibitem{Schneider20130246}
Schneider CM, Belik V, Couronn{\'e} T, Smoreda Z, Gonz{\'a}lez MC.
\newblock Unravelling daily human mobility motifs.
\newblock Journal of The Royal Society Interface. 2013;10(84).
\newblock doi:{10.1098/rsif.2013.0246}.

\bibitem{Montjoye13}
de~Montjoye YA, Hidalgo CA, Verleysen M, Blondel VD.
\newblock Unique in the Crowd: The privacy bounds of human mobility.
\newblock Scientific Reports. 2013;3.

\bibitem{gonzalez08}
Gonzalez MC, Hidalgo CA, Barabasi AL.
\newblock Understanding individual human mobility patterns.
\newblock Nature. 2008;453(7196):779--782.

\bibitem{Bassolas2016}
Bassolas A, Lenormand M, Tugores A, Gon{\c{c}}alves B, Ramasco JJ.
\newblock Touristic site attractiveness seen through Twitter.
\newblock EPJ Data Science. 2016;5(1):12.
\newblock doi:{10.1140/epjds/s13688-016-0073-5}.

\bibitem{Jurdak2015}
Jurdak R, Zhao K, Liu J, AbouJaoude M, Cameron M, Newth D.
\newblock Understanding Human Mobility from Twitter.
\newblock PLOS ONE. 2015;10(7):1--16.
\newblock doi:{10.1371/journal.pone.0131469}.

\bibitem{FriasMartinez2014a}
Frias-Martinez V, Frias-Martinez E.
\newblock Spectral clustering for sensing urban land use using Twitter
  activity.
\newblock Engineering Applications of Artificial Intelligence. 2014;35:237 --
  245.
\newblock doi:{http://doi.org/10.1016/j.engappai.2014.06.019}.

\bibitem{Arase2010}
Arase Y, Xie X, Hara T, Nishio S.
\newblock Mining People's Trips from Large Scale Geo-tagged Photos.
\newblock In: Proceedings of the 18th ACM International Conference on
  Multimedia. MM '10. New York, NY, USA: ACM; 2010. p. 133--142.
\newblock Available from: \url{http://doi.acm.org/10.1145/1873951.1873971}.

\bibitem{Kisilevich2010}
Kisilevich S, Mansmann F, Keim D.
\newblock P-DBSCAN: A Density Based Clustering Algorithm for Exploration and
  Analysis of Attractive Areas Using Collections of Geo-tagged Photos.
\newblock In: Proceedings of the 1st International Conference and Exhibition on
  Computing for Geospatial Research and Application. COM.Geo '10. New York, NY,
  USA: ACM; 2010. p. 38:1--38:4.
\newblock Available from: \url{http://doi.acm.org/10.1145/1823854.1823897}.

\bibitem{Giannotti2007}
Giannotti F, Nanni M, Pinelli F, Pedreschi D.
\newblock Trajectory Pattern Mining.
\newblock In: Proceedings of the 13th ACM SIGKDD International Conference on
  Knowledge Discovery and Data Mining. KDD '07. New York, NY, USA: ACM; 2007.
  p. 330--339.
\newblock Available from: \url{http://doi.acm.org/10.1145/1281192.1281230}.

\bibitem{Tracy}
Farrell TA, Marion JL.
\newblock Trail Impacts and Trail Impact Management Related to Visitation at
  Torres del Paine National Park, Chile.
\newblock Leisure/Loisir. 2001;26(1-2):31--59.

\bibitem{Arrowsmith2002295}
Arrowsmith C, Inbakaran R.
\newblock Estimating environmental resiliency for the Grampians National Park,
  Victoria, Australia: a quantitative approach.
\newblock Tourism Management. 2002;23(3):295 -- 309.

\bibitem{Shoval2010}
Shoval N.
\newblock Monitoring and Managing Visitors Flows in Destinations using
  Aggregative {GPS} Data.
\newblock In: Information and Communication Technologies in Tourism, {ENTER}
  2010, Proceedings of the International Conference in Lugano, Switzerland,
  February 10-12, 2010; 2010. p. 171--183.
\newblock Available from: \url{http://dx.doi.org/10.1007/978-3-211-99407-8_15}.

\bibitem{Strogatz2001}
Strogatz SH.
\newblock Exploring complex networks.
\newblock Nature. 2001;410(6825):268--276.
\newblock doi:{http://dx.doi.org/10.1038/35065725}.

\bibitem{Barthelemy2011}
Barthélemy M.
\newblock Spatial networks.
\newblock Physics Reports. 2011;499(1-3):1--101.
\newblock doi:{10.1016/j.physrep.2010.11.002}.

\bibitem{watts1998cds}
Watts DJ, Strogatz SH.
\newblock {Collective dynamics of'small-world'networks.}
\newblock Nature. 1998;393(6684):409--10.

\bibitem{Latora2002109}
Latora V, Marchiori M.
\newblock Is the Boston subway a small-world network?
\newblock Physica A: Statistical Mechanics and its Applications.
  2002;314(1-4):109 -- 113.
\newblock doi:{http://dx.doi.org/10.1016/S0378-4371(02)01089-0}.

\bibitem{Poland22}
Sienkiewicz J, Ho\l{}yst JA.
\newblock Statistical analysis of 22 public transport networks in Poland.
\newblock Phys Rev E. 2005;72:046127.
\newblock doi:{10.1103/PhysRevE.72.046127}.

\bibitem{Zhang2016}
Zhang X, Chen G, Han Y, Gao M.
\newblock Modeling and Analysis of Bus Weighted Complex Network in Qingdao City
  Based on Dynamic Travel Time.
\newblock Multimedia Tools Appl. 2016;75(24):17553--17572.
\newblock doi:{10.1007/s11042-016-3376-4}.

\bibitem{Soh2010}
Soh H, Lim S, Zhang T, Fu X, Lee GKK, Hung TGG, et~al.
\newblock Weighted complex network analysis of travel routes on the Singapore
  public transportation system.
\newblock Physica A: Statistical Mechanics and its Applications.
  2010;389(24):5852--5863.

\bibitem{TIAN2016537}
Tian Z, Jia L, Dong H, Su F, Zhang Z.
\newblock Analysis of Urban Road Traffic Network Based on Complex Network.
\newblock Procedia Engineering. 2016;137:537 -- 546.

\bibitem{Knights13}
Knights AM, Koss RS, Robinson LA.
\newblock Identifying common pressure pathways from a complex network of human
  activities to support ecosystem-based management.
\newblock Ecol Appl. 2013;23(4):755--65.

\bibitem{Coombes2008}
Coombes EG, Jones AP, Bateman IJ, Tratalos JA, Gill JA, Showler DA, et~al.
\newblock Spatial and Temporal Modeling of Beach Use: A Case Study of East
  Anglia, UK.
\newblock Coastal Management. 2009;37(1):94--115.
\newblock doi:{10.1080/08920750802527127}.

\bibitem{Douglas73}
Algorithms for the reduction of the number of points required to represent a
  digitized line or its caricature.
\newblock Cartographica: The International Journal for Geographic Information
  and Geovisualization. 1973;10(2):112--122.
\newblock doi:{10.3138/fm57-6770-u75u-7727}.

\bibitem{Friedman1977}
Friedman JH, Bentley JL, Finkel RA.
\newblock An Algorithm for Finding Best Matches in Logarithmic Expected Time.
\newblock ACM Trans Math Softw. 1977;3(3):209--226.
\newblock doi:{10.1145/355744.355745}.

\bibitem{VlachosGK02}
Vlachos M, Gunopulos D, Kollios G.
\newblock Discovering Similar Multidimensional Trajectories.
\newblock In: Agrawal R, Dittrich KR, editors. ICDE. IEEE Computer Society;
  2002. p. 673--684.
\newblock Available from:
  \url{http://dblp.uni-trier.de/db/conf/icde/icde2002.html#VlachosGK02}.

\bibitem{refAlgorithm}
Lera I. MatchGPX: a python algorithm to detect overlapped regions between a
  pair of GPX trajectories; 2016.
\newblock \url{https://github.com/wisaaco/MatchGPX}.

\bibitem{Comaniciu02meanshift}
Comaniciu D, Meer P.
\newblock Mean shift: A robust approach toward feature space analysis.
\newblock IEEE Transactions on Pattern Analysis and Machine Intelligence.
  2002;24:603--619.

\bibitem{refWikiloc}
Wikiloc;.
\newblock Available from: \url{https://www.wikiloc.com/}.

\bibitem{refDataSet}
Lera I. Dataset of Human Hiking GPS Trajectories on Balearic Islands (SPAIN).;
  2017.
\newblock \url{https://osf.io/77n9t}.

\bibitem{wunderground}
Weather Underground;.
\newblock Available from: \url{https://www.wunderground.com/}.

\bibitem{hagberg2008}
Hagberg AA, Schult DA, Swart PJ.
\newblock Exploring network structure, dynamics, and function using {NetworkX}.
\newblock In: Proceedings of the 7th Python in Science Conference (SciPy2008).
  Pasadena, CA USA; 2008. p. 11--15.

\bibitem{LouvainMethod}
Blondel VD, Guillaume JL, Lambiotte R, Lefebvre E.
\newblock Fast unfolding of communities in large networks.
\newblock Journal of Statistical Mechanics: Theory and Experiment.
  2008;2008(10):P10008.

\bibitem{ICWSM09154}
Bastian M, Heymann S, Jacomy M. Gephi: An Open Source Software for Exploring
  and Manipulating Networks; 2009.
\newblock Available from:
  \url{http://www.aaai.org/ocs/index.php/ICWSM/09/paper/view/154}.

\bibitem{Szabo2004}
Szab{\'o} G, Alava M, Kert{\'e}sz J.
\newblock In: Ben-Naim E, Frauenfelder H, Toroczkai Z, editors. Clustering in
  Complex Networks. Berlin, Heidelberg: Springer Berlin Heidelberg; 2004. p.
  139--162.
\newblock Available from: \url{http://dx.doi.org/10.1007/978-3-540-44485-5_7}.

\bibitem{Sienkiewicz05}
Sienkiewicz J, Ho\l{}yst JA.
\newblock Statistical analysis of 22 public transport networks in Poland.
\newblock Phys Rev E. 2005;72:046127.
\newblock doi:{10.1103/PhysRevE.72.046127}.

\bibitem{Chowell03}
Chowell G, Hyman JM, Eubank S, Castillo-Chavez C.
\newblock Scaling laws for the movement of people between locations in a large
  city.
\newblock Phys Rev E. 2003;68:066102.
\newblock doi:{10.1103/PhysRevE.68.066102}.

\bibitem{Newman}
Newman MEJ.
\newblock Assortative Mixing in Networks.
\newblock Phys Rev Lett. 2002;89:208701.
\newblock doi:{10.1103/PhysRevLett.89.208701}.

\bibitem{Li2007}
Li W, Cai X.
\newblock Empirical analysis of a scale-free railway network in China.
\newblock Physica A: Statistical Mechanics and its Applications.
  2007;382(2):693--703.

\bibitem{Amaral2000}
Amaral LA, Scala A, Barthelemy M, Stanley HE.
\newblock Classes of small-world networks.
\newblock Proc Natl Acad Sci U S A. 2000;97(21):11149--11152.
\newblock doi:{10.1073/pnas.200327197}.

\bibitem{Moreira2010}
Moreira A, Santos MY, Wachowicz M, Orellana D.
\newblock In: Painho M, Santos MY, Pundt H, editors. The Impact of Data Quality
  in the Context of Pedestrian Movement Analysis. Berlin, Heidelberg: Springer
  Berlin Heidelberg; 2010. p. 61--78.
\newblock Available from: \url{http://dx.doi.org/10.1007/978-3-642-12326-9_4}.

\bibitem{Brakatsoulas2005}
Brakatsoulas S, Pfoser D, Salas R, Wenk C.
\newblock On Map-matching Vehicle Tracking Data.
\newblock In: Proceedings of the 31st International Conference on Very Large
  Data Bases. VLDB '05. VLDB Endowment; 2005. p. 853--864.
\newblock Available from:
  \url{http://dl.acm.org/citation.cfm?id=1083592.1083691}.

\bibitem{Shang2014}
Shang S, Ding R, Zheng K, Jensen CS, Kalnis P, Zhou X.
\newblock Personalized trajectory matching in spatial networks.
\newblock The VLDB Journal. 2014;23(3):449--468.
\newblock doi:{10.1007/s00778-013-0331-0}.

\end{thebibliography}

\end{document}